\documentclass[12pt]{article}
\usepackage{amsmath}
\usepackage{xcolor}
\definecolor{DarkBlue}{rgb}{0,0,0.5451}
\definecolor{DarkGreen}{rgb}{0,0.39216,0}

\usepackage[pdftex,plainpages=false,pdfpagelabels,colorlinks=true,pdfpagemode=UseOutlines,citecolor=DarkBlue,%
	urlcolor=DarkGreen%
]{hyperref}
\usepackage[pdftex]{graphicx}
\usepackage{enumerate}
\usepackage{natbib}
\usepackage{url}

\newcommand{\blind}{0}
\newcommand{\elearn}{e-learning platform}
\newcommand{\R}{\textsf{R}}

\begin{document}
\graphicspath{{./figures/}}
\DeclareGraphicsExtensions{.pdf,.png}

\def\spacingset#1{\renewcommand{\baselinestretch}%
{#1}\small\normalsize} \spacingset{1}


\if0\blind
{
  \title{\bf More than Formulas - Integrity, Communication, Computing and Reproducibility in Statistics Education}
  \author{Eva Furrer, 0000-0002-1997-9744\thanks{
    The authors gratefully acknowledge Reto Gerber, Institute of Molecular Life Science, University of Zurich, for his help in the initial setup of the CI/CD functionalities of the course's Gitlab repository. Eva Furrer was partially supported by the P-8 ‘Digital Skills’ in Teaching 2021-2024 project of the University of Zurich. }\hspace{.2cm}\\
    Epidemiology, Biostatistics and Prevention Institute,\\
    Center for Reproducible Science \\
    and \\
    Annina Cincera \\
    Epidemiology, Biostatistics and Prevention Institute,\\ Department of Mathematical Modeling and Machine Learning\\    
    and \\
    Reinhard Furrer, 0000-0002-6319-2332 \\
    Department of Mathematical Modeling and Machine Learning,\\ University of Zurich}
  \maketitle
} \fi

\if1\blind
{\LARGE\vspace*{2cm}
  \begin{center}
    {\bf More than Formulas - Integrity, Communication, Computing and Reproducibility in Statistics Education\\[1cm]}
\end{center}
  \medskip
} \fi

\begin{abstract}
This paper introduces a novel course design in the Master Program in Biostatistics at the University of Zurich that integrates computing skills, effective communication, reproducibility, and scientific integrity within one course. Utilizing a flipped classroom model, the course aims to equip students with the necessary competencies to handle real-world data analysis challenges and effective statistical practice in general. The curriculum includes practical tools such as version control with Git, dynamic reporting, unit testing and containerization to foster reproducibility, and integrity in statistical practice. Feedback gathered from both staff and students post-implementation indicates that the course significantly enhances student readiness for professional and academic environments, demonstrating the effectiveness of this educational approach.

\end{abstract}

\noindent%
\vfill

\newpage
\spacingset{1.45} 
\section{Introduction}
\label{sec:intro}

With the growing importance of statistics and data science in research and society at large, attention to statistics and data science education has been growing concurrently. \citet{doi:10.1080/00031305.2016.1270230}, for example, describe the principles behind the design of a master program in applied statistics. Among other insights, the authors emphasize that a skilled statistician must possess a set of component skills beyond mathematical understanding, including computing, communication, authentic practice, and critical thinking skills. Using a survey among alumni, the authors conclude that such skills are just as important for job preparedness as the methodological instruction provided by the program. \citet{NIKIFORIDOU2010795} widen this view to instruction in statistics more generally and an appeal to move away from traditional mathematical focus towards reasoning and interpretation.

More than ten years ago \citet{doi:10.1198/tast.2010.09132} already called for a reform of statistics curricula towards providing much more computational and programming competencies. \citet{doi:10.1080/10691898.2020.1870416} summarize the developments in the intervening decade and list many useful resources. In the same special issue  \citet{doi:10.1080/10691898.2020.1804497} describe their effort to provide a course addressing computational and statistical thinking for an audience without requiring a mathematical background, whereas \citet{doi:10.1080/10691898.2020.1845109} inform on ways to provide software engineering skills to students of statistics.

\citet{doi:10.1080/00031305.1991.10475825} state that ``the interdisciplinary role of a statistician demands that one be able to communicate statistical hypotheses, procedures, and results to a variety of audiences at various levels.'' Indeed, communication skills are critical for statisticians but are rarely included in statistics and data science curricula \citep{doi:10.1080/26939169.2022.2138800}. For data science, \citet{doi:10.52041/serj.v21i2.40} advocate a more `holistic approach to data science education'' including instruction on how to communicate of key findings. 

Paying sufficient attention to reproducibility and workflow is part of the authentic practice skills statisticians and data scientists should possess. Several examples are given in the editorial by \citet{doi:10.1080/26939169.2022.2141001}: ``Robust workflows matter. For instance, COVID-19 counts in the United Kingdom were underestimated because the way that Excel was used resulted in dropped data [\dots].'' The editorial introduces eleven invited papers on “Teaching reproducibility and responsible workflow.”  \citet{doi:10.1080/26939169.2022.2074922} claim that the additional challenge in teaching how to work reproducibly is that often students are less motivated and most of the tools have a steep learning curve, to overcome this challenge they advise to placing extra emphasis on motivation, using guided instruction and allowing time for lots of practice.
Many more recommendations for incorporating reproducibility concepts into teaching in statistics and data science are provided by \citet{doi:10.1080/26939169.2022.2138645}. Also, for introductory statistics courses, i.e., not necessarily in a statistics or data science curriculum, \citet{doi:10.5070/T581020118} propose using R Markdown to avoid an antiquated copy-paste workflow.

Version control is an important part of making workflows more reproducible and, therefore, is also part of the required skill set of statisticians and data scientists. The widely used distributed version control system Git and the associated developer platforms GitHub and GitLab originated in software engineering but are today also standard in a research and education environment. The use of Git is notoriously difficult to learn and to integrate into a personal working routine, see, e.g., \citet{doi:10.1080/10691898.2019.1617089}, and \citep{doi:10.1080/10691898.2020.1848485} in a statistics and data science education context. Nevertheless, \citet{doi:10.1080/10691898.2019.1617089} promote its use and describe how using Git decreases the administrative work for staff in a class. The article provides useful resources to set up courses using GitHub Classroom. \citep{doi:10.1080/10691898.2020.1848485} present a wide range of approaches to teaching Git at different levels and backgrounds, serving as an inspiration in many settings. Finally, as mentioned by several authors \citep{doi:10.1080/00031305.2016.1270230, NIKIFORIDOU2010795, doi:10.1080/26939169.2022.2074922} working on real projects is important for the development of appropriate component skills and project work is easily implemented in a Git environment. Note that data privacy issues may make using the commercial GitHub server undesirable \citep{doi:10.1080/10691898.2020.1848485}; institutional GitLab platforms are one possible solution for this issue.

We can, therefore, conclude that much has already been said on the topic of introducing computing, communication, authentic practice, and critical thinking skills into statistics or data science curricula. Why do we think another article describing another teaching initiative needs to be written and disseminated? With this article, we aim to provide a very concrete example of teaching good practices to students in a statistics or close to statistics program, combining all of the separate topics above into one course.

A flipped classroom model for designing a course in the described context seems appropriate for several reasons. First, several studies have shown that the outcome of introductory statistics courses, measured in grades and student satisfaction, is better in the flipped classroom, 
\citet{doi:10.52041/serj.v17i1.179} and 
\citet{doi:10.1177/0098628315620063} show this in a quasi-experimental approach, \citet{doi:10.1177/0098628313487461} and \citet{doi:10691898.2014.11889717} using historical controls. Second, the very successful model of participatory live coding of the global nonprofit The Carpentries (\url{https://carpentries.org}) promises to be a very effective activity type for the synchronous course parts \citep{Wils:16}. \citet{doi:journal.pcbi.1006023} and \citet{doi:10.1371/journal.pcbi.1008090} provide very useful tips on how to teach programming in general and how to best set up participatory live coding. Third, \citet{doi:journal.pcbi.1006023} and \citet{doi:10.1080/10691898.2020.1848485} discuss peer instruction as being a useful activity in such technical courses and the Git system lends itself to an easy implementation of such an activity in asynchronous course parts. 

This resulting flipped classroom course is easily adaptable to other contexts and settings, and we provide pertinent material to make the setup of a similar offer very easy, including templates to exploit continuous integration capacities via GitLab. 

\section{Motivation for the course}
\label{sec:why}
Our motivation for introducing the proposed course in the curriculum of the Master Program in Biostatistics is similar to the discussed articles of Section~\ref{sec:intro}. Data are ubiquitous in today's society and even more so in research. A very large part of research projects, hence, rely on obtaining insight from data that are specifically collected for the project come from existing databases, or are scraped from the web. Statisticians should be involved in the entire research process of such projects, from project design via data collection, processing, and analysis to interpreting and discussing the gained insights. Their contribution, besides core statistical expertise, includes rigorous checks for the integrity and reliability of the data, an emphasis on reproducibility and robustness of the data analysis pipeline, and in-depth guidelines on communicating the results, see, for example, \citet{10.1371/journal.pcbi.1004961} for a summary why this is effective statistical practice. Our experience in the Master Program in Biostatistics at the University of Zurich (\url{www.biostat.uzh.ch}) since 2011 is that students do not acquire the relevant skills for such a role simply as a byproduct of their statistics lectures and required research projects. A good statistics curriculum needs to include specific preparation for this role. 

Starting with pilot courses in 2018, a single dedicated course called ``Good Statistical Practice'' was introduced in Fall 2021. It spans four fundamental topics: integrity, communication, computing, and reproducibility. 

\section{Course content and structure}
\label{sec:what}
The course's core audience are biostatistics students and students of the Minor Program in Applied Probability and Statistics who major in diverse subjects, e.g., biology, psychology, economics, banking, finance, etc. All of these students have had detailed introductions to statistics and have already gained experience in using \R.

\begin{figure}
\centering
\includegraphics[width=.7\textwidth]{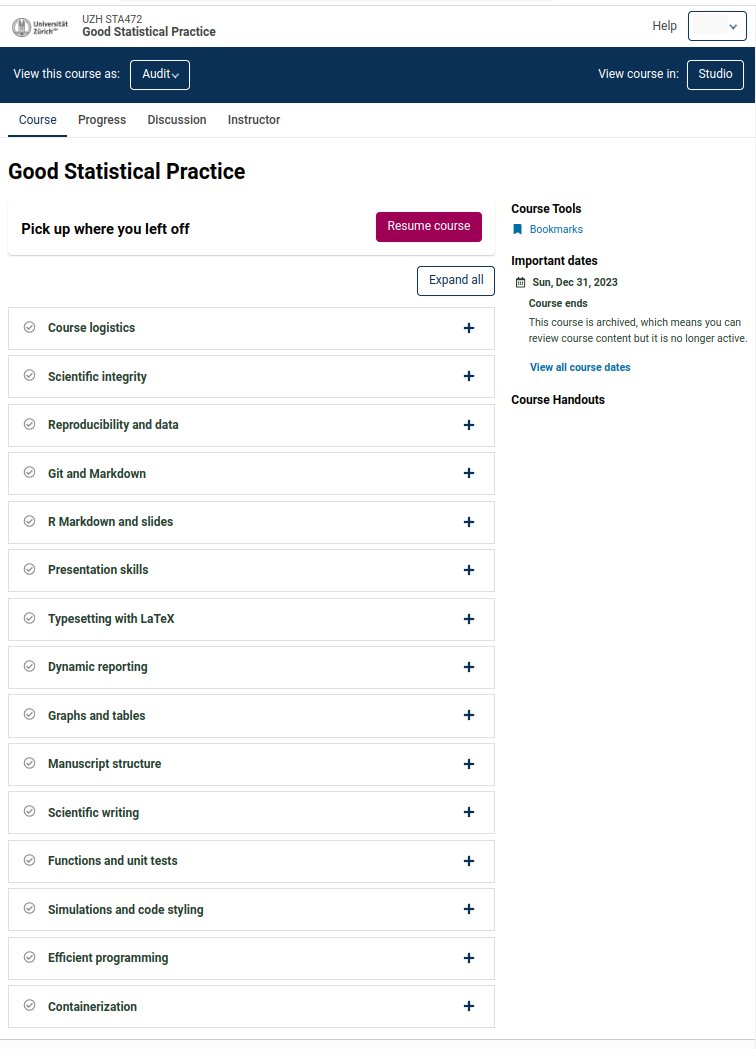}
\caption{Content and breakdown into the chapters of the course on the \elearnƒelearn. \label{fig:coursecontent}}
\end{figure}

The lecture period at the University of Zurich is 14 weeks. Consequently, the course was structured into 14 chapters; Figure~\ref {fig:coursecontent} indicates the chapter titles.

\begin{figure}
\centering
\includegraphics[width=\textwidth]{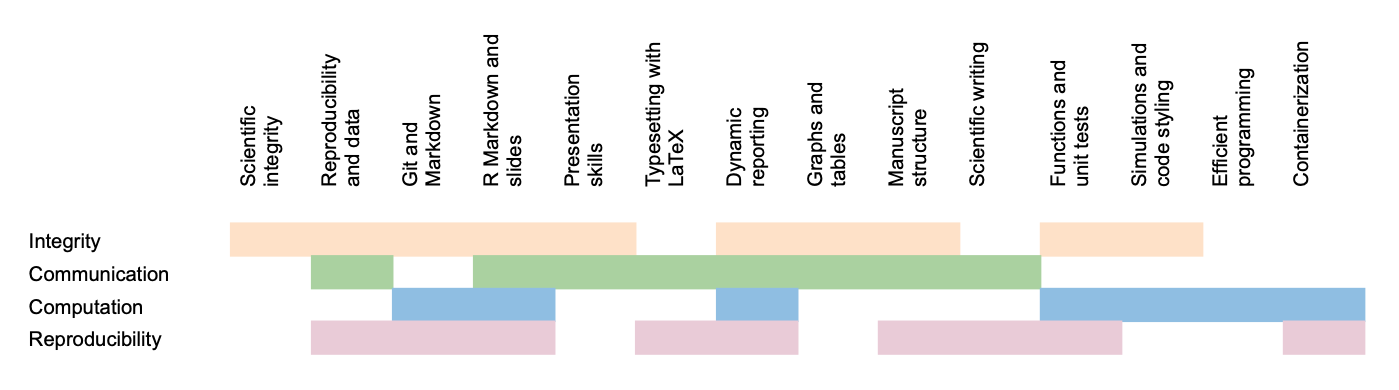}
\caption{Matrix design of the course with the four fundamental topics coverage throughout the 14 chapters. \label{fig:gantt}}
\end{figure}

The red thread through almost all of the chapters is a semi-real data example from microbiology that does not require deep biological insight to understand a few simple research questions. The fundamental topics of integrity, communication, computing, and reproducibility are recurring themes while students work on the example. The matrix design of the course representing the four topics throughout the 14 weeks is shown in Figure~\ref{fig:gantt}. Communication skills, specifically presentation and writing skills, are taught in a continuous block in the middle of the course period, building on the computational skills and focusing on reproducibility covered beforehand. The course concludes with a focus on more advanced computational and reproducibility themes. The topic of scientific integrity recurs during the 14 weeks whenever reliability and accountability are the motivation for requiring students to acquire a specific skill.

\section{Teaching method}
\label{sec:how}
\begin{figure}
\centering
\includegraphics[width=.7\textwidth]{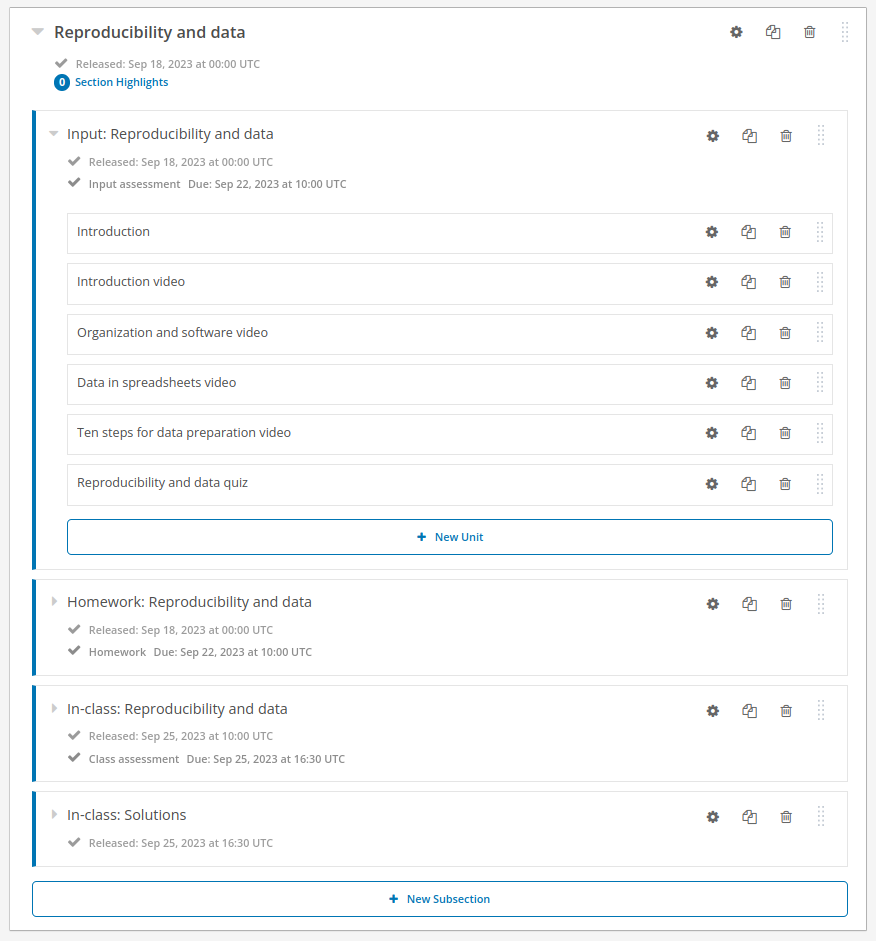}
\caption{Example of chapter structure in \elearn. \label{fig:chapterstructure}}
\end{figure}

\begin{figure}
\centering
\includegraphics[width=.43\textwidth]{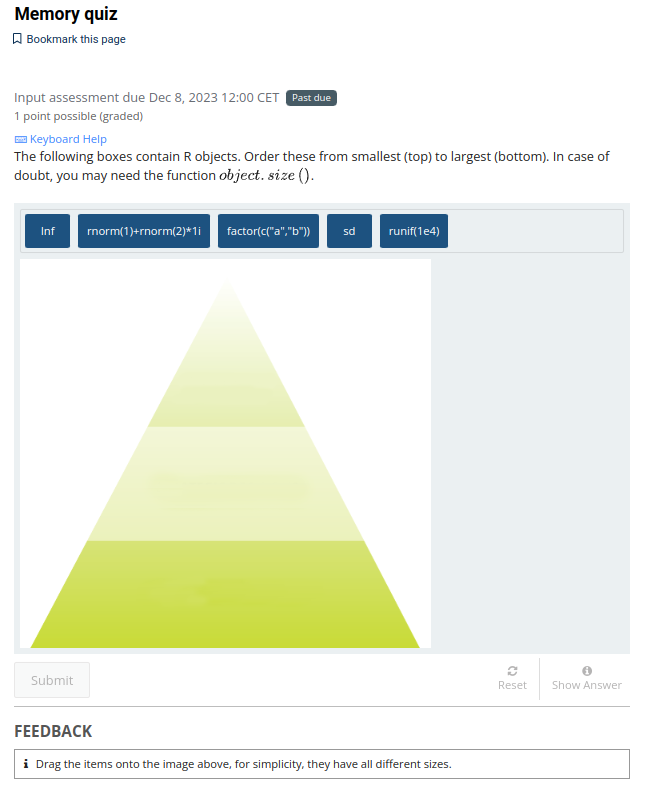}\hfill
\includegraphics[width=.55\textwidth]{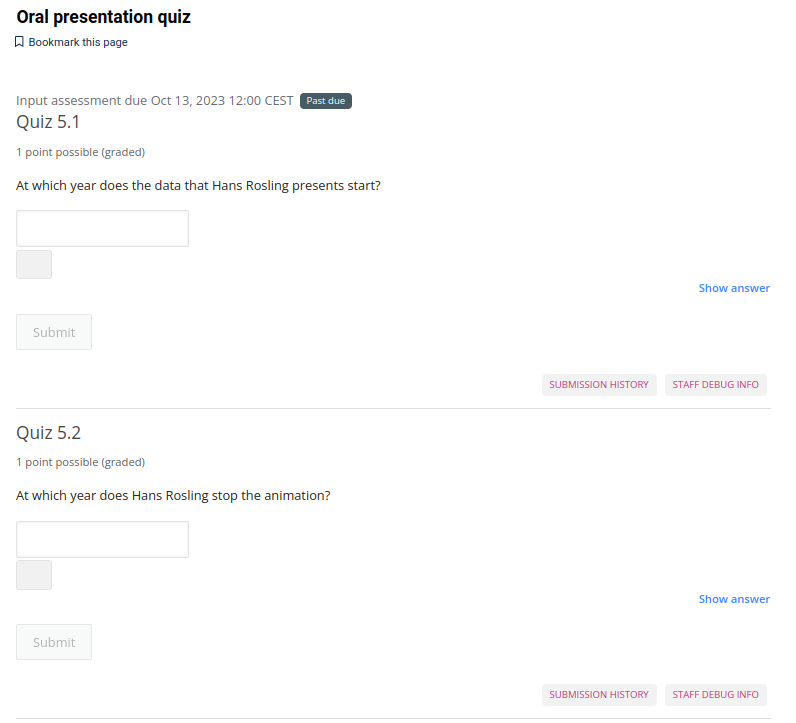}\\
\caption{Examples of quizzes in \elearn. \label{fig:quizzes}}
\end{figure}

The general design of the course is a flipped learning setup realized on an institutional instance of Open edX (\url{openedx.org}). Each chapter is structured into asynchronous input, homework, and a synchronous in-class section. See, for example, the chapter ``Reproducibility and data'' in Figure~\ref{fig:chapterstructure}. The input typically consists of an introduction video followed by a sequence of videos about detailed content with intermediate quizzes. Figure~\ref{fig:quizzes} indicates some of the types of quizzes that are used. 

\begin{figure}
\centering
\includegraphics[width=.7\textwidth]{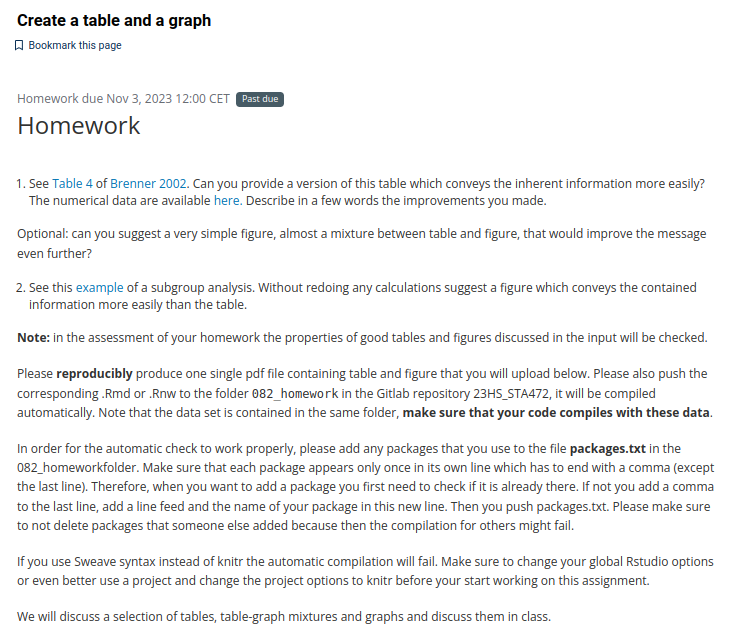}
\caption{Example of a homework in \elearn. \label{fig:homework}}
\end{figure}

Following the input, students must work on a homework task that requires them to apply the concepts they learned in the input in a concrete situation. Figure~\ref{fig:homework} shows an example from the chapter ``Graphs and Tables''. Part of the homework is to peer-review the work of one to three other students and, with that, learn to critically assess the type of results required in the homework. Staff spot-checks the peer reviews and, for some chapters, provides feedback directly.

\begin{figure}
\centering
\includegraphics[width=.7\textwidth]{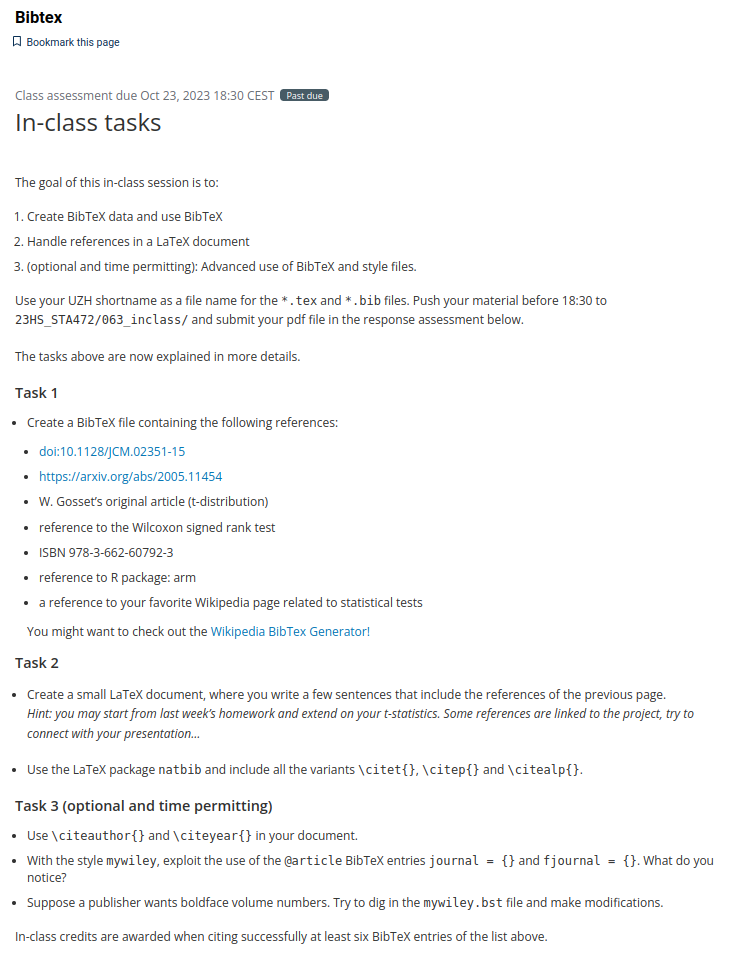}
\caption{Example of an in-class task in \elearn. \label{fig:in-class}}
\end{figure}

Finally, during the in-class session, problems occurring during the homework session are discussed. Students who did not get useful feedback from peer review of their homework can ask for better help. The input and homework provide a solid basis for students to work on a substantial task during the in-class time. The focus of these sessions is to give practical help for those tasks, e.g., through carpentries-style live-coding instructions. Figure~\ref{fig:in-class} shows an example from the chapter ``Typesetting with LaTeX''.

One recurrent theme of the course is that students must submit their homework on the University of Zurich institutional GitLab server; staff then checks compilability through a continuous integration setup. The goal of this requirement is for students to practice using version control throughout the semester so that it becomes a habit for them in the future. For the staff, the setup also has the advantage of using a generated list of student results indicating directly for whom problems with the compilation have occurred.

Besides version control, this element of repetition is also used for dynamic reporting and, of course, the four fundamental topics of integrity, communication, computing, and reproducibility. 

All results that students provide, i.e., input quizzes, homework with peer review, and in-class tasks, are accounted for on the Open edX platform with points. This portfolio examination replaces a traditional exam. Students pass if they gain at least 70\% of the total points; there are no grades for the course.

A two-software setting (Open edX and Gitlab) was used, which is conceptually sub-optimal, as neither directly communicates nor exchanges information with the other. Some aspects of the Open edX functionality could have been implemented using GitHub Classroom or addressed with a \texttt{ghclass} \citep{ghclass} framework, but not everything is possible (see the specific sections in \citealp{doi:10.1080/10691898.2019.1617089, doi:10.1080/10691898.2020.1848485}). The two-software setting was adopted as a pragmatic compromise.

\section{Lessons Learned}
\label{sec:lesson}
The course has been offered in the three fall semesters, 2021, 2022, and 2023. In fall 2023, it was made compulsory in the Master Program in Biostatistics curriculum. 

\subsection{Staff feedback}
From the point of view of the perception of the teaching staff of the course and the master program, the new course has, at least for a large part of the cohorts, succeeded in
\begin{itemize}
    \item making students use version control and dynamic reporting as their default way of operating
    \item improving the quality of presentations and reports in the project-based course ``Statistical Consulting''
    \item preparing students for many of the challenges of their master thesis
    \item providing students with a good starting skill set for more involved programming tasks in their future
    \item shaping students' habits towards prioritizing integrity, reproducibility, and diligence in data analysis
\end{itemize}

\subsection{Student survey}
Besides the perception of the teaching staff, we have also conducted a short survey of the three student cohorts. We could only reach those students who still maintain a University of Zurich mail account or had provided us with alternative means of contacting them. In total, 68 persons booked the course throughout the three fall semesters. From these persons 60 university email addresses were still functioning, out of the 8 for which the university email address did not work anymore we could contact 4 through other addresses. 22 persons participated in the survey. This corresponds to a response rate of 32\% of the total cohort and 34\%  of the cohort of students who were successfully contacted.

Out of the 22 respondents, 9 participated in Fall 2021, 5 in Fall 2022, and 8 in Fall 2023; 14 were students of the Master Program in Biostatistics, 6 of the Minor Program in Applied Probability and Statistics, and 2 from other programs. Twelve respondents reported that they used the learned concepts only in an academic setting, and 10 reported using them also in a non-academic setting.

Regarding the content of the course, we asked students the following questions regarding the concepts below:
\begin{enumerate}
    \item On a scale from never since, rarely, sometimes, often to frequently, did you use the following concepts or tools since you took the course?
   \item If you used the following concepts at least rarely, on a scale from not really, somewhat,  to definitively, did the training of the course help you?
\end{enumerate}

The concepts covered were ``Good practice for"
spreadsheets, file and folder organization, version control, dynamic reporting, \LaTeX, presentation slide design, oral presentations, designing graphs, designing tables, structure for manuscript, logic of a paragraph, writing style, writing \R\ functions, using unit tests, setting up simulations, code styling, writing vectorized code, writing parallelized code, containerizing code.

The options ``I do not know" and ``I do not use this concept" were also available for the second question. See the Supplementary Material for screenshots of how these questions were delivered and how we obtained students' consent to use the provided answers.

\begin{figure}
\includegraphics[height=.411\textwidth]{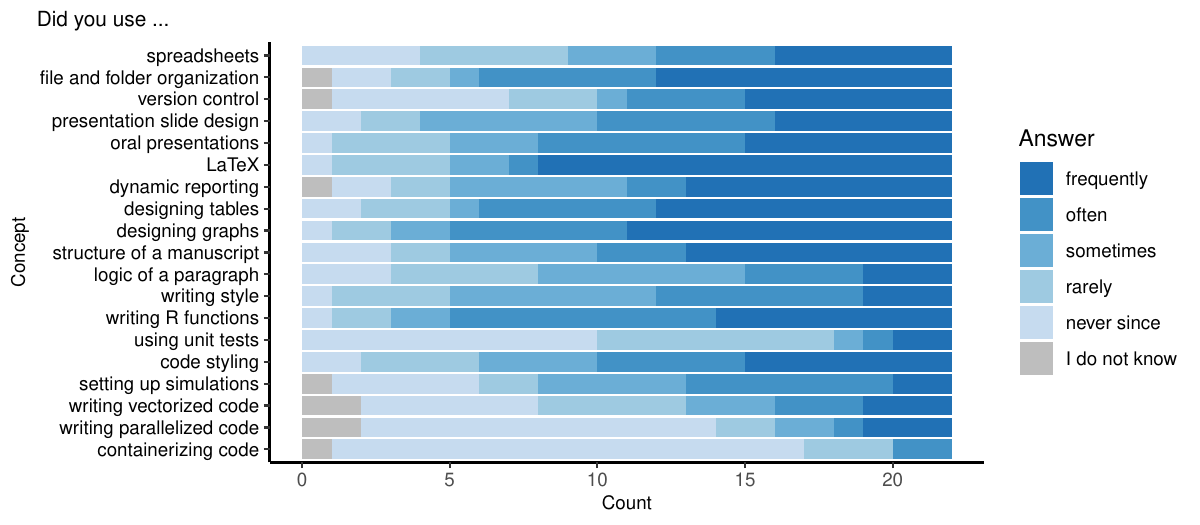}
\includegraphics[height=.411\textwidth]{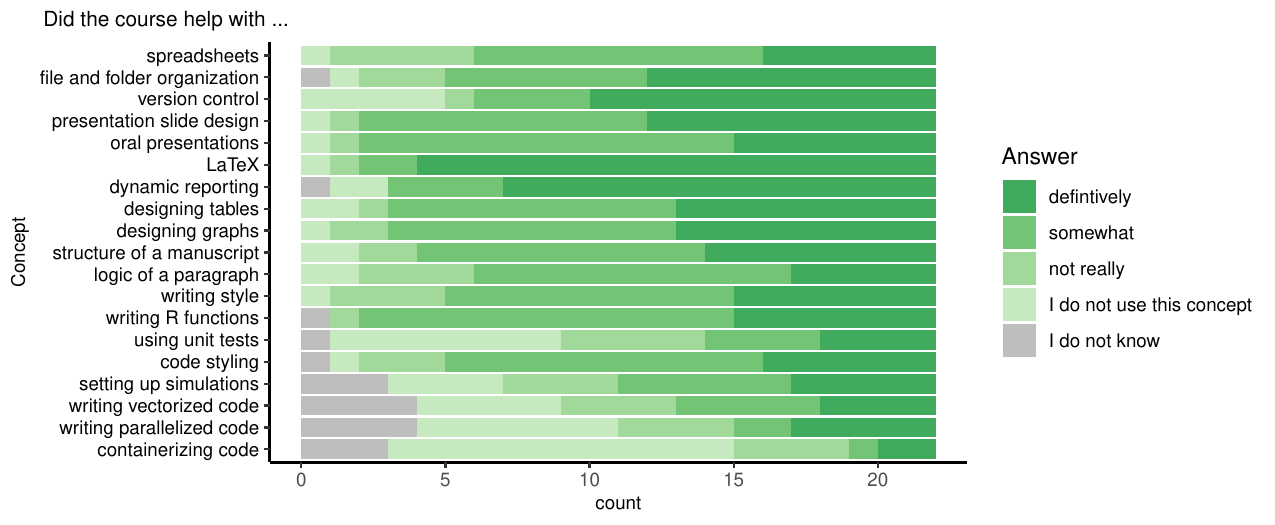}
\caption{The distribution of the answers of the survey. \label{fig:survey}}
\end{figure}

The distribution of the answers by the 22 respondents is given in Figure~\ref{fig:survey}. Overall, many of the respondents used many of the learned concepts after the course and found the course helpful for almost all concepts. Without surprise we observe that the more advanced topics have been less used and the course helped less than for other topics. 

\section{Discussion}
\label{sec:discussion}
The introduction of the course ``Good Statistical Practice" has been perceived as important and successful by the teaching staff of the Master Program in Biostatistics and the participating students, see Section~\ref{sec:lesson}. The course has been set up to use the teaching staff's time efficiently; hence, these benefits are not associated with high costs. Should student numbers increase, resources would naturally need to be increased as well; this is specifically true for the course's presentation and writing skills parts. While adaptations to the writing skills part, which include the responsible exploitation of generative large language model implementations, are already in development, it is unclear how a scale-up for the presentation skills part could be possible.

With increasing understanding and mastery of version control, students quickly realized that they could always push their assignments to Gitlab even if the deadline on the Open edX system had passed. As a solution, students were required 
to submit a corresponding Git hash of their work. However, in the end, it was not tested to see if the actual hash existed and contained the entire assignment since only very few infractions were observed. Students were also never observed back-dating commits to extend deadlines. Moreover, the open setup, allowing students to see peers' submissions, was necessary for the peer review design of the course. A negative consequence was that students could easily copy someone else's work as their own. Such (rare) occasions were used to demonstrate that the transparency of version control allows one to completely see through such situations.

The content of the individual topics is tailored for a reasonable learning curve: the principle ``Start Slowly and Keep It Simple'' \citep{doi:10.1080/10691898.2020.1848485} was followed in the course design. Consequently, not all possible technical aspects are covered, e.g., sensitive data and data protection in general, advanced macro writing and custom formatting hooks in \LaTeX, rebasing and submodules in Git, load balancing and resource management in \R\, etc.

Finally, it is worth mentioning that we introduced a similar module through the School for Transdisciplinary Studies of the University of Zurich, i.e., addressing all students no matter their discipline (under the condition of basic statistics and \R\ knowledge) and exposing them to an abbreviated and less technical version of the course.

\section{Ethics and data availability statement}
The University of Zurich provides an assessment tool that helps determine whether a study involving human subjects needs to be reviewed by an ethics committee under Swiss Federal or Zurich cantonal law. The assessment for this article's student survey confirmed that no review by an ethics committee is necessary. 

The supplementary material for this article contains the consent form that survey participants signed.

The data supporting this study's findings are openly available in Zenodo at \url{https://doi.org/10.5281/zenodo.12703467}.

\bibliographystyle{Chicago}

\bibliography{bibliography}
\end{document}